# Unitary response of solvatochromic dye to pulse excitation in lipid and cell membranes

Simon Fabiunke[1]*, Christian Fillafer[1], Matthias F. Schneider[1]*

[1]Department of Medical and Biological Physics, Technical University Dortmund; 44227 Dortmund, Germany.

*Corresponding authors.

Simon Fabiunke: Otto-Hahn-Straße 4, 44227 Dortmund, Germany, +49 231 755 2994, simon.fabiunke@tu-dortmund.de.

Matthias F. Schneider: Otto-Hahn-Straße 4, 44227 Dortmund, Germany, +49 231 755 4139, matthias-f.schneider@tu-dortmund.de.

**Abstract**

The existence of acoustic pulse propagation in lipid monolayers at the air-water interface is well known. These pulses are controlled by the thermodynamic state of the lipid membrane. Nevertheless, the role of acoustic pulses for intra- and intercellular communication are still a matter of debate. Herein, we used the dye di-4-ANEPPDHQ, which is known to be sensitive to the physical state and transmembrane potential of membranes, in order to gain insight into compression waves in lipid-based membrane interfaces. The dye was incorporated into lipid monolayers made of phosphatidylserine or phosphatidylcholine at the air-water-interface. A significant blue shift of the emission spectrum was detected when the state of the monolayer was changed from the liquid expanded (LE) to the liquid condensed (LC) phase. This "transition-sensitivity" of di-4-ANEPPDHQ was generalized in experiments with the bulk solvent dimethyl sulfoxide (DMSO). Upon crystallization of solvent, the emission spectrum also underwent a blue shift. During compression pulses in lipid monolayers, a significant fluorescence response was only observed when in the main transition regime. The optical signature of these waves – in terms of sign and magnitude – was identical to the response of di-4-ANEPPDHQ during action potentials in neurons and excitable plant cells. These findings corroborated the suggestion that action potentials are nonlinear state changes that propagate in the cell membrane.

**Main Text**

**Introduction**

The Langmuir technique is a common tool to record static state diagrams of lipid monolayers which represent one leaflet of a bilayer membrane[1]. The recorded state diagrams show a strong thermodynamic coupling between several observables such as area per molecule (A), lateral pressure ($\pi$), temperature (T), surface potential ($\psi$) etc.[2] These couplings have also been observed during acoustic pulses that were excited in the lipid-based interface[3–6]. There is an ongoing debate if such dynamic state changes play a major role in biological communication. It has been shown for example that the activity of enzymes embedded into lipid membranes[4] as well as the fluorescent emission of dye molecules[5] are strongly coupled to the thermodynamic state of the interface. This implies that propagating pulses may transiently activate/deactivate biological membrane functions.

Many studies of biological systems have relied on the use of fluorescent molecules. The light that is emitted from fluorophores can help to localize certain structures/molecules of interest in a given sample. Aside from spatial clues, it is known that the emitted light can also carry information about the physical properties of the system in which the fluorophore is embedded[7–10].

We and others have characterized the optical response of di-4-ANEPPDHQ, which is a member of the ANEPP-family of dyes. ANEPP-dyes have been used as "transmembrane potential sensors" in neurophysiological research. Some studies, however, have demonstrated that the specificity of the dye response for transmembrane potential changes is by no means absolute. It was shown, for example, that ANEPP-dyes also respond to membrane fluidity and/or dipole orientation changes[11]. Other studies have indicated that the fluorescence emission from di-4-ANEPPDHQ can be used to distinguish ordered from disordered membrane domains[7,12–14]. In previous studies it was found that the emission spectrum of this dye indeed undergoes a significant shift when a membrane interface becomes ordered at the main transition[7,13,15]. This indicates that the dye is rather sensitive to the thermodynamic state of its solvation shell, i.e. it behaves like solvatochromic dyes such as LAURDAN[9,16,17]. During a transition the thermodynamic susceptibilities of a membrane like the compressibility, heat capacity or electrical capacity exhibit a maximum which means that the system is highly sensitive to perturbations (temperature, pressure, pH, etc.). When di-4-ANEPPDHQ was incorporated into an excitable plant cell membrane (*Chara* internode), nonlinear shifts of the emission spectrum were found in the temperature- and pH-state diagrams. In particular, the emission spectrum blue-shifted upon cooling and acidification, *i.e.,* upon two environmental changes which typically lead to ordering of lipid membranes[15]. Interestingly, a transient blue-shift was also observed when an action potential (AP) was triggered in the cell. Taken together, these results indicated that an excitable cell membrane resides close to a transition. Furthermore, the membrane interface undergoes a transient ordering during propagation of an AP, which is in line with predictions of Kaufmann[18], Heimburg[19] and others[20–22].

While previous studies by us and others had already indicated that di-4-ANEPPDHQ is sensitive to the phase state of a lipid membrane, there was still a lack of data about the optical response throughout a transition regime. Herein, we filled this gap by recording the emission spectrum of di-4-ANEPPDHQ over a range of phase states in lipid membranes and DMSO. For this purpose, protein-free monolayers made of phosphatidylserine (PS) and phosphatidylcholine (PC) were employed. In all lipid membranes as well as in DMSO, the emission spectrum of the dye underwent a sigmoidal blue shift upon ordering at the main transition. This invariance of the optical response with respect to the type of solvent clearly demonstrated the solvatochromic mechanism of the spectral shift of di-4-ANEPPDHQ. Furthermore, we – for the first time – used this dye to track the propagation of density pulses in lipid monolayers. The emission spectrum responded with a transient shift during a pulse, but only when the membrane was perturbated in the main transition regime. The optical response in the lipid monolayer was identical to the ones that had been recorded during action potentials in excitable plant cells and neuronal tissue. This unitary response of the solvatochromic dye di-4-ANEPPDHQ to pulses in living and non-living systems indicates that action potentials are nonlinear state changes that propagate in the cell membrane interface.

**Materials and Methods**

**Materials** Di-4-ANEPPDHQ was purchased from Thermo Fisher Scientific and dissolved in chloroform ($c = 1$ mg/ml). The two phospholipids, 1,2-dimyristoyl-sn-glycero-3-phospho-L-serine (DMPS) dissolved in chloroform:methanol:water (65:35:8) and 1,2-dipalmitoyl-sn-glycero-3-phosphocholine (DPPC) dissolved in chloroform were obtained from Avanti Polar Lipids. All other reagents were of analytic purity and were purchased from Sigma Aldrich.

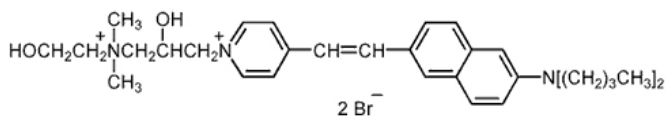

Figure 1: Chemical structure of di-4-ANEPPDHQ.

**Opto-mechanical characterization of fluorescent monolayer** A Langmuir trough was used to opto-mechanically characterize the lipid monolayers with embedded di-4-ANEPPDHQ. The maximal system area was 275 $cm^2$ and the minimal area was 75 $cm^2$. Two barriers made of Teflon coupled to a motor were used to change the area with a velocity of 10 $cm^2/min$. The lateral pressure of the lipid monolayer was detected by a Wilhelmy-plate method. To form a lipid monolayer, a lipid solution ($c \approx 1$ mg/ml) containing fluorophore of up to 4 mol% was carefully spread at the air-water interface until an increase of lateral pressure was detected. The subphase contained an aqueous buffered salt solution (100 mM NaCl, 10 mM phosphate buffer, pH 7) at room temperature (~20 °C).

The emission spectrum of the fluorescent dye was recorded by means of a fluorescence microscope that was mounted on top of the film balance. The dye was excited by a high-power LED with a maximal intensity at 470 nm. The emitted photons were reflected by a dichromate (transmission > 520 nm, reflection < 500 nm) to either a spectrum analyzer (Wasatch Photonics WP-00183, spectral range: 400 nm to 1080 nm) or to two photomultiplier tubes equipped with different band pass filters ($(610 \pm 10)$ nm and $(690 \pm 10)$ nm). Before each measurement a background signal of the water surface was recorded which was subtracted from all data that were recorded subsequently. Since fluorescence emission from the monolayer was not stable at large film areas (low lateral pressure) over an extended period of time, the layer was initially formed at small area. If the signal was lost, an additional droplet of the lipid dye solution (V~1-2 µL) was spread. Furthermore, the system was allowed to equilibrate for five minutes after monolayer formation to exclude effects of residual solvent molecules on the emission spectrum.

**Excitation and opto-mechanical detection of density pulses** A lateral density perturbation was excited by local exposure of the monolayer to hexane vapor (s. Figure S1). For this purpose, a stainless-steel cylinder was filled with nitrogen gas at an absolute pressure of 1.6 bar. Upon opening of a valve, the compressed gas streamed into the gas phase of a bottle which was partially filled with hexane. The gas stream carried a fraction of the evaporated solvent onto the monolayer. When hexane dissolves into the lipid-based interface, it locally perturbated the system. As a consequence of conservation of momentum, this perturbation propagated through the layer as a compression front.

The lateral pulses in the lipid monolayer were detected mechanically by a pressure sensor (Wilhelmy method) and optically by a ratiometric setup (two photomultipliers with bandpass filters at 610 nm and 690 nm; readout rate: 4 kHz). Since the propagation velocity of the waves and the distance between the sensors was known, the mechanical and optical signal of a pulse can be correlated. To ensure that a pulse indeed propagated across the optical field, the pressure sensor was placed some centimeters behind the optical detector. To get a stable optical signal, it was necessary to reduce the motion of the surface of the monolayer. For this purpose, additional barriers made of Teflon were placed into the Langmuir trough (**SI, Figure S1**). A single measurement cycle started at small lateral pressures. By titrating small amounts of the lipid-dye mixture onto the air-water interface the lateral pressure was increased and the phase state of the lipid monolayer was controlled. Acoustic pulses were excited in different regimes across a typical state diagram.

**Spectral shift during in bulk solution** To observe the emission spectrum of di-ANEPPDHQ (1 µM) dissolved in DMSO as a function of temperature, an optically transparent chamber was set on top of a Peltier element. The Peltier element was used to vary the environmental temperature from 30 °C to 5 °C. At every temperature step the system was allowed to equilibrate for 5 min. The emission spectrum of the solution was detected with a spectrum analyzer (Wasatch Photonics, WP-00183, spectral range: 400 nm to 1080 nm).

**Results and Discussion**

**Optical signature of acoustic pulses in a lipid-based interfaces**. One of the routine techniques of neurophysiology is to track cell membrane excitation with potential-sensitive fluorophores. These molecules are incorporated into cellular membranes and subsequently the dye response is monitored during external stimulation or spontaneous cellular activity. Herein, we proceeded in an analogous manner. We incorporated the dye di-4-ANEPPDHQ in a protein-free lipid membrane at an air-water-interface. It was demonstrated previously that adiabatic pulses can be excited by various means in such a system[3,20,23,24]. The different manifestations of acoustic pulses in lipid monolayers have been detected with mechanical[3,6,23], electrical[6], and optical techniques[3,25,26]. Herein, we used an opto-mechanical Langmuir trough setup (s. SI, Figure S1) to study the fluorescence response of di-4-ANEPPDHQ, a member of the ANEPP-family of dyes, to acoustic pulses in lipid monolayers.

**Isothermal state changes**. In order to understand fluorescence emission changes in lipid monolayers, we first conducted quasi-stationary (isothermal) experiments. The ratio parameter $r$ was recorded while a PS monolayer was compressed by a motorized barrier. The ratio parameter r relates the emission intensities in two bands ($610 \pm 10$ nm and $690 \pm 10$ nm) of the emission spectrum to each other ($r = I_{610}/I_{690}$). Therefore, changes of the ratio parameter indicate shifts of the emission spectrum.

The pressure-area isotherm of the fluorescent monolayer displayed a nonlinearity at the main transition where the monolayer changes from the liquid-expanded (LE) to the liquid-condensed (LC) phase (**Figure 2a, inset**). The midpoint of the transition is defined as the pressure where the derivative of the state diagram, *i.e.* the compressibility, has a maximum (transition pressure $\pi$ =17.9 m/mN, **Figure 2a**). In **Figure 2b** the optical susceptibility ($\eta_T = \frac{1}{r}\frac{dr}{d\pi}$; response of the ratio parameter to changes in lateral pressure) is shown. The optical susceptibility $\eta_T$ was maximal in the transition regime, where it was 10 times larger as compared to the LE and LC phases. These results confirmed previous findings of a blue shift of the emission spectrum of di-4-ANEPPDHQ when a lipid membrane is taken across the main transition[7,12,15]. A qualitatively identical optical response was found when the experiments were conducted with a lipid monolayer made of a different lipid (phosphatidylcholine) (s. SI, Figure S2, S3). This indicated that the significant spectral shift across the main transition is invariant with respect to the headgroup of the phospholipid.

**Adiabatic state changes**. Density pulses were excited in three fundamentally different phase state regimes: liquid-expanded phase **1** (LE phase), liquid-condensed phase **3** (LC phase) and near the transition regime **2** (s. **Figure 2a**). The monolayer resting state was perturbed by locally blowing hexane gas onto the interface. This resulted in a pulse-like change of lateral pressure akin to a compression wave. The amplitude of these pulses was in the range of ~1.5 mN/m and only varied slightly in the different regimes of the state diagram (s. SI, Figure S4). The response of *r* to these compression pulses is shown in **Figure 2c**. The noise of the optical signal was about 2 %. To compare the changes during acoustic pulses the ratio parameter was normalized at the start of each measurement. When a density pulse was excited in the LE **1** or LC phase **3**, no significant changes of r were observed ($< 2\,\%$). In some experiments, additional small variations of the baseline signal (with no correlation to the lateral pressure pulse) appeared. These were probably artifacts which can be caused by convection or capillary waves at the air-water interface.

An entirely different optical response was obtained when a compression pulse was excited at the LE-edge **2** of the main transition (**Figure 2c**). Although the pressure amplitude did not change significantly between the measurements (s. SI, Figure S4), the ratio parameter increased by $19 \pm 2$ % (n =5). This was indicative of a transient shift of the emission spectrum to shorter wavelengths.

Based on the isothermal characterization, we concluded that the pulse transiently ordered the system and took it into the transition regime, because there the spectral shift is largest.
only when the system was perturbed into a transition regime.

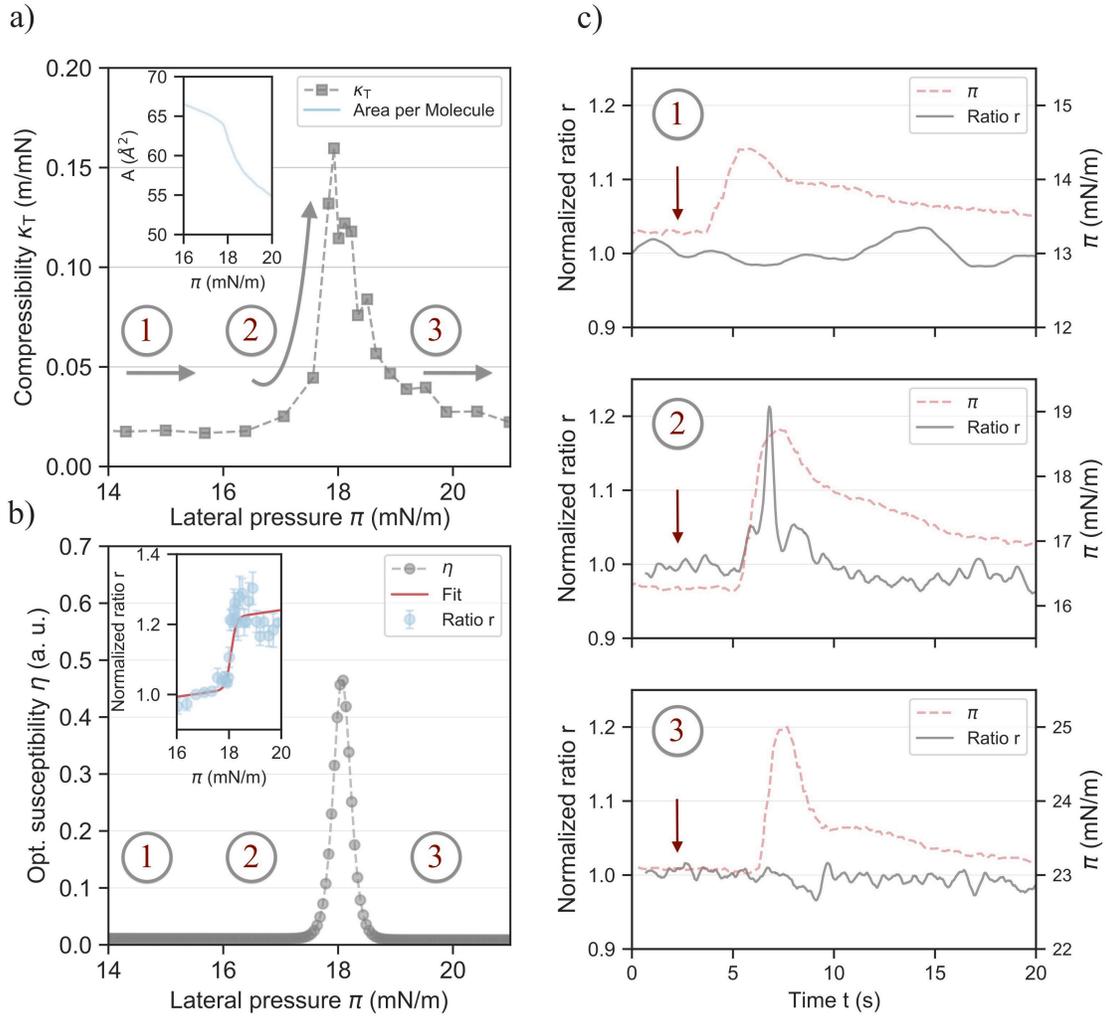

**Figure 2: State-dependent optical response of di-4-ANEPPDHQ in lipid monolayer. (a)** The isothermal compressibility $\kappa_T$ in the three different regimes for the pulse experiments in a DMPS monolayer (liquid expanded (LE) **1**, liquid-condensed (LC) **3**, and at the LE-edge of the main transition **2** respectively). The excitation of density pulses in the regimes **1**, **2** and **3** is indicated by the grey arrows. The inset of **(a)** shows the area per molecule as a function of lateral pressure. The vertical line indicates the transition pressure where the compressibility is maximal. **(b)** The optical susceptibility $\eta_T$ and the ratio parameter $r$ (inset) as a function of lateral pressure $\pi$. In the transition regime the response of the optical signal to quasi-static compression is about 10 times higher than outside the transition. **(c)** Exemplary responses of $\pi$ and $r$ to compression pulses (red arrow indicates time point of excitation). There was only a significant response in state **2**, namely when the resting state was at the LE-edge of the transition regime (total number of experiments: n=5; for additional data sets see Figure S5).

**Opto-thermal characterization of fluorescent solution.** In order to investigate if the spectral response of di-4-ANEPPDHQ depends on mechanical changes or on location of the dye at a macroscopic interface, spectrometric experiments were conducted in absence of such an interface, *i.e.,* with fluorophore in bulk solution. For this purpose, the dye was dissolved in dimethyl sulfoxide (DMSO) at a concentration of 1 μM at 30 °C. This temperature is above the melting point of DMSO at atmospheric pressure ($T_m$ = 18 °C[27]). The system was then cooled and the emission spectrum was recorded. The ratio parameter changed in a sigmoidal manner upon freezing (s. **Figure 3**). As in the lipid monolayer, ordering at the main transition was associated with a spectral shift and therefore an increase of the ratio parameter *r*. The change of *r* was ~40% and thus of the same order of magnitude as in the lipid monolayer experiments. The optical susceptibility $\eta_p$ (derivative of a sigmoidal fit of *r* as a function of temperature) was maximal at 17 °C which is in excellent agreement with the melting point of DMSO[27]. Since the melting point is associated with a maximum of the heat capacity of the system, this demonstrated that optical and thermal susceptibilities of the system were coupled and both became maximal at the transition. Such a coupling was also observed in the lipid monolayer experiments (s. **Figure 2** or [25]), where it extends to other susceptibilities such as the electrical capacity, membrane compressibility and thermal expansion coefficient[2]. The coupling of different susceptibilities demonstrates a central point, namely that a change of state is in general associated with changes of many (if not all) observables of the system. The magnitude of these changes can be extracted from the phenomenology, *i.e.* from the respective state diagrams. In this regard the derivatives of the state diagrams (susceptibilities) are particularly important. The susceptibilities become maximal in a transition which means that state changes in a transition are associated with maximal changes of the extensive observables. In other words, when a system is close to a transition, small changes in temperature, pressure or surface potential lead to large changes in enthalpy and the emission spectrum of the dye.

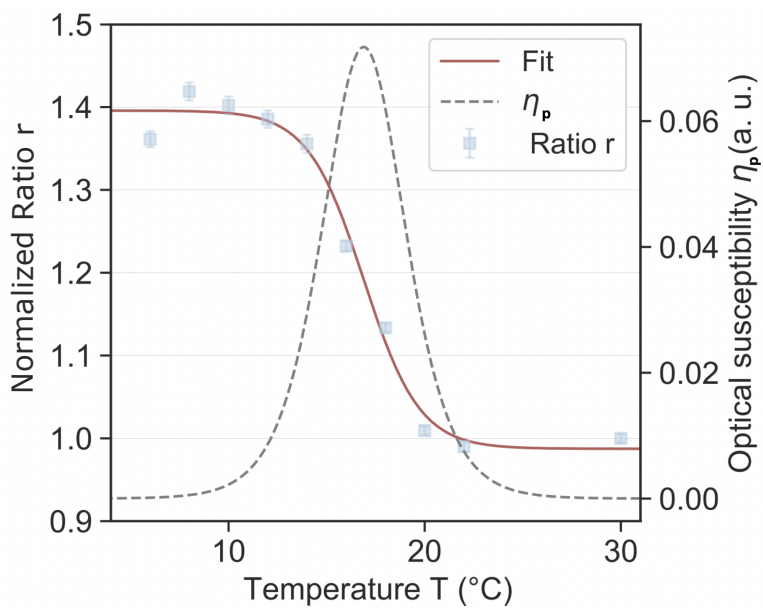

**Figure 3: State-dependent fluorescence emission of Di-4-ANEPPDHQ in dimethyl sulfoxide (DMSO)**. The ratio parameter *r* as a function of temperature. Each data point represents the ratio of the fluorescence intensities in two bands (600 nm-620 nm and 680 nm - 700 nm) of the emission spectrum. The error bars result from the error propagation of the intensity uncertainty in the emission bands. The optical susceptibility $\eta_p$ ($\propto \frac{dr}{dT}$), i.e. the maximal change of the ratio r with temperature, is given by the derivative of a sigmoidal fit. Its maximum at $T_m = 17$ °C coincides with the melting point of DMSO at atmospheric pressure ($T_{m,\text{DMSO}} = 18$ °C[27]). Note that the optical susceptibility $\eta_p$ is about 10 times larger in the transition regime.

**Comparison of optical signature in lipid monolayer and excitable cells.** ANEPP-dyes have been employed as sensors of transmembrane potential[10,28] and as sensors of lipid ordering[12,14,29,30]. Herein, it was shown that emission from di-4-ANEPPDHQ is clearly sensitive to isothermal state changes and pressure pulse perturbations of a lipid monolayer at the air-water interface. The emission spectrum of the dye also underwent a significant shift at the melting transition of a bulk solvent (DMSO). In both of these experimental systems Nernst potentials are absent by definition. This is in line with the findings of others, who have shown that ANEPP-dyes are sensitive to fluidity[7] and dipole orientational changes[11]. Therefore, a broad basis of evidence now corroborates that the emission spectrum of the dye has a solvatochromic response. We believe that the common interpretation of ANEPP-dye signal changes in (neuro-)biology as Nernst potential changes has to be revisited in light of these data.

The coupling between thermodynamic variables calls for careful use of the term "specific response". If the coupling is tacitly ignored, one might be inclined to call di-4-ANEPPDHQ an electrosensitive[10,22], mechanosensitive (c.f. **Figure 2**) or thermosensitive probe (c.f. **Figure 3**) respectively. However, all conclusions are limited because they originate from the viewpoint of the observer, who conducted a "mechanical", "electrical" and "thermal" experiment respectively. When the different viewpoints are combined, it becomes apparent that the change of the emission spectrum is maximal in the transition regime and the latter can be reached by various combinations of thermodynamic quantities in phase space (thermal, mechanical, chemical, electrical, etc.).

It was demonstrated herein that the emission spectrum of di-4-ANEPPDHQ shifts to shorter wavelengths when the system undergoes an ordering transition. The spectral shift and therefore magnitude of the optical response was largest in the transition regime (**Figure 2** and **3**). Furthermore, it was shown that density pulses in a lipid membrane are only associated with a significant optical response in vicinity of a transition (**Figure 2c**). These results imply that whenever a significant shift of the emission spectrum of di-4-ANEPPDHQ is detected, one has to first and foremost suspect that a state change close to a transition is involved. Di-4-ANEPPDHQ and closely related fluorophores have been used previously for studies of pulse propagation in living systems[15,28]. When APs were triggered in neurohypophyses and excitable plant cells, a unitary fluorescence response was observed (**Figure 4 a), b)**). Intriguingly, this response was identical with the optical signature of an acoustic pulse in the main transition regime of a monolayer (**Figure 4 c)**). In all of the three systems - *neurohypophysis, excitable plant cell and lipid monolayer* – the propagating pulse was associated with a reversible increase of the ratio parameter. This suggests a blue shift of the emission spectrum and therefore a transient ordering of the membrane interface. The simplest explanation for this finding is that pulses close to a transition in lipid monolayers and APs in excitable cell membranes are the same phenomena caused by the same natural laws. This conclusion implies that neither Nernst potentials nor specific proteins should have an *a priori* status in the interpretation of the fluorescence signal during APs. While both components may have a role in setting the thermodynamic state of an excitable membrane, they cannot be the natural cause of pulse propagation in a lipid membrane[18,32].

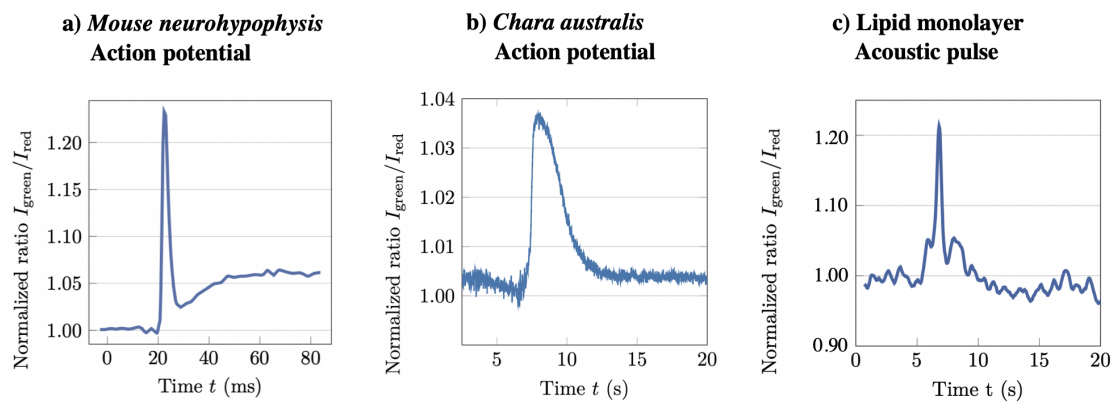

**Figure 4: Optical signatures of action potentials and acoustic pulses close to a transition.**
(a) Optical signal of the related fluorescent dye RH 414 upon excitation of mouse neurohypophysis (data taken from (20)). (b) Optical signal of di-4-ANEPPDHQ upon triggering an action potential in excitable plant cells (*Chara australis,* Figure taken from[15]). (c) Response of di-4-ANEPPDHQ to perturbation of a lipid monolayer close to a transition. In all three systems a very similar change of the ratio parameter occurred during pulse propagation.

**Conclusion**

The present experiments revealed a thermodynamic coupling of the emission spectrum of di-4-ANEPPDHQ to the phase state of a system. In lipid membrane as well as bulk solvent (DMSO), ordering at the main transition was associated with a blue shift of the spectrum. It was demonstrated that the optical response can be used to track the propagation of compression waves in lipid membranes. When the pulse forced the system state into the main transition, a characteristic optical signature was observed. Intriguingly, the same response had been recorded during action potentials in excitable cells. This corroborated that an AP is a nonlinear acoustic pulse which propagates in the cell membrane because of conservation laws[18,20,22,33–36]. Such a pulse must manifest itself in changes of all observables of the membrane, one being the electrical potential and another being the emission spectrum of an embedded fluorophore.

**Acknowledgments**

The authors thank K. Kaufmann, S. Shrivastava and L. Bagatolli for valuable discussions.

**Author Contributions:**
M.F.S. designed and supervised the overall project. S.F. and C.F. designed the experiments. S.F. performed the experiments and did the experimental analysis. All authors wrote and reviewed the main manuscript.
**Competing Interest Statement:** The authors declare that they have no competing interests.

**Data availability Statement:** The authors confirm that the data supporting the findings of this study are available within the article and its supplementary materials.

## Optical detection of acoustic pulses

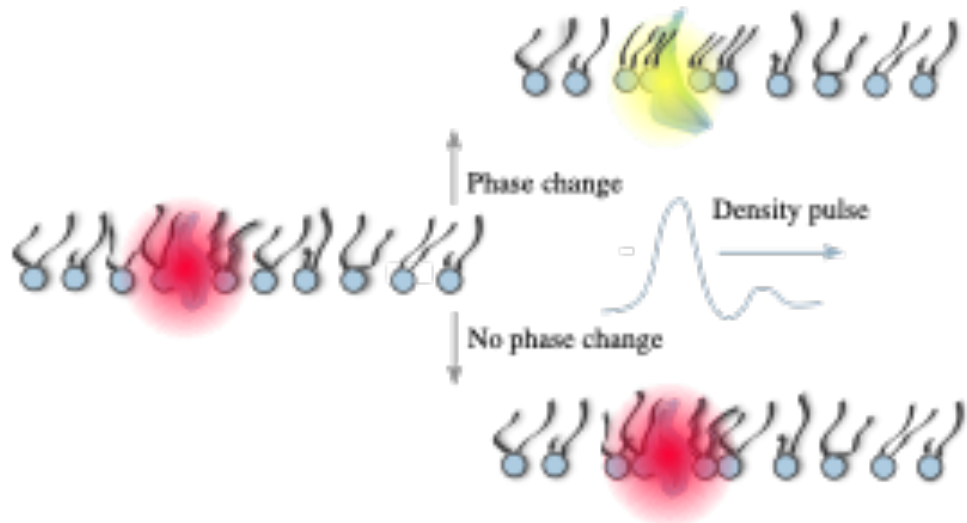